# Interfacial Thermal Conductance Between a Polyethylene Glycol Polymer Chain and Water: A Molecular Dynamics Study

Shadi Babaei[1], Yekta Cheraghali[1], Claire Loison[2], Ali Rajabpour[1*], Samy Merabia[2*]

1. Mechanical Engineering Department, Imam Khomeini International University, Qazvin, Iran
2. Univ Lyon, Univ Claude Bernard Lyon 1, CNRS, Institut Lumière Matière, F-69622, Villeurbanne, France
   * correspondance: a.rajabpour@gmail.com, samy.merabia@univ-lyon1.fr

## Abstract

Understanding interfacial heat transfer between polymers and water is crucial for the design of biomaterials, drug delivery platforms, and nano-fluidic systems. In this study, we employed all-atom molecular dynamics (MD) simulations to quantify the interfacial thermal conductance between a polyethylene glycol (PEG) 36-mer chain and explicit water over the temperature range of 280–350 K. To compare the conformational behavior of the PEG chain, we examined its radius of gyration and observed a temperature-dependent chain collapse consistent with previous coarse-grained models. By employing a transient non-equilibrium MD approach, we imposed temperature difference across the interface and analyzed the energy relaxation behavior to compute heat transfer across the polymer-water interfaces. Our results demonstrate that both temperature and interfacial interaction strength influence interfacial thermal conductance, with temperature playing the dominant role. Structural factors such as chain conformation and interfacial area were found to mediate the effect of interfacial interaction. Additional analysis of the vibrational density of states (VDOS) and the mean square displacement (MSD) reveal that vibrational coupling has minimal impact on thermal conductance across interfaces, whereas increased water thermal motion enhances energy transfer. These findings highlight the structural and dynamical origins of interfacial thermal conductance and provide atomistic insights into the tuning of interfacial heat transport in molecular systems through temperature and solvent interactions.

# 1. Introduction

The thermal transport properties of polymers have been the subject of an increasing number of investigations driven by thermal management issues [1-5]. Among polymers, understanding the thermal transport properties of macromolecules in aqueous environments is crucial for applications in biomaterials, drug delivery systems, and nanofluidics [1,2]. Following the development of scanning force microscopes, the thermal transport properties of molecules have been characterized at the single molecule level [6,7] opening the door to investigate thermoelectricity of polymers [8]. Macromolecules have been also traditionally employed as capping agents to regulate thermal transport between metallic nanoparticles and their aqueous environment [9-11].

Energy transport across macromolecules in water has been characterized by spectroscopy techniques [12-17], atomistic simulations [18,19] and theoretical approaches [20]. Thermal relaxation in these systems is controlled by the competition between inelastic energy scattering inside the molecule and heat transfer to the aqueous environment [21-23]. However, the influence of polymer conformation on energy transport has been scarcely addressed in the literature. To fill this gap, water soluble polymers are interesting candidates as their conformation may change under the action of temperature.

Among water soluble polymers, polyethylene glycol (PEG) is particularly notable due to its biocompatibility, tunable solubility, and widespread use in biomedical engineering [24, 25]. A key factor governing PEG's functionality in these applications is its interfacial thermal coupling with water, which controls heat dissipation in polymer-based nanodevices, drug release kinetics, and thermoresponsive materials [26].

The hydrophilic nature of PEG facilitates hydrogen bonding with water, leading to temperature-dependent solvation dynamics and conformational flexibility [27]. Previous studies have explored PEG's structural behavior in aqueous solutions: Lee demonstrated that low-molecular-weight PEG adopts an ideal chain conformation in water, with corresponding scaling exponents consistent with experimental observations [28]. Later, Chudoba et al. developed a coarse-grained PEG model valid across a wide range of temperature (270–450 K), accurately predicting its radius of gyration and osmotic pressure [29]. However, despite these advances, quantitative insights into the thermal transfer mechanisms at the PEG-water interface remain limited.

Recent studies on biomolecule-water interfaces provide valuable benchmarks for such investigations [30]. For instance, Hamzi et al. reported interfacial thermal conductance (ITC) values of 40–80 $MW \cdot m^{-2} \cdot K^{-1}$ for amino acids in water, highlighting the role of hydrophobicity and molecular structure in heat transfer [31]. Similarly, Lervik and Kurisaki examined separately thermal transport at protein-water Interfaces, finding ITC values ranging from 100–270 $MW \cdot m^{-2} \cdot K^{-1}$, with structural flexibility and hydration dynamics playing key roles [32, 33]. These findings suggest that polymer-water ITC is highly sensitive to molecular conformation and interfacial bonding, yet systematic studies on PEG are lacking.

In this work, we employ molecular dynamics (MD) simulations to quantify the interfacial thermal conductance between a PEG chain and water. We examine how factors such as chain length, temperature, and interaction strength influence interfacial heat transfer. In particular, we investigate how the chain configuration – collapsed versus extended – impacts thermal transfer to water. The results are interpreted in terms of phonon mode analysis and mobility of the water molecules. By bridging the gap between polymer physics and nanoscale thermal transport, our study provides fundamental insights for designing PEG-based materials with optimized thermal properties for biomedical and nanotechnological applications.

## 2. Simulation Method

In the present study, molecular dynamics (MD) simulations were conducted using the LAMMPS package [34] to investigate the thermal behavior and energy exchange mechanisms between a polyethylene glycol (PEG) chain and surrounding water molecules.

The initial configuration of the system was generated using the Polymer Builder module of the CHARMM-GUI interface [35]. A single PEG chain consisting of 36 monomer units (36-mer) was constructed and solvated in a cubic water box (100 Å per side) with periodic boundary conditions in all directions as represented in figure 1. The CHARMM36 force field was used to define both bonded and non-bonded interactions, and the TIP3P model was employed to represent water molecules, consistently with the CHARMM-GUI Polymer builder interface.

Bonded interactions—including bonds, angles, dihedrals, and impropers were described using CHARMM force field parameters [36]. Non-bonded interactions between PEG and water molecules were modeled using Lennard-Jones (LJ) and Coulombic terms. The LJ potential employed here has the standard 12-6 form:

$$U(r) = 4\chi\varepsilon \left[ (\sigma/r)^{12} - (\sigma/r)^{6} \right] \qquad (1)$$

where $\varepsilon$ is the energy well depth, $\sigma$ is the molecule diameter, $r$ is the interatomic distance, and $\chi$ represents the van der Waals (vdW) dimensionless coupling strength between PEG and water molecules. The LJ parameters for the cross interactions were determined using the Lorentz mixing rule.

The simulation protocol consisted of three main stages. First, the system underwent energy minimization followed by thermal equilibration under an NPT ensemble for 1 ns using a Nosé–Hoover thermostat and barostat at temperatures ranging from 280-350 K. Next, a short 0.5 ps simulation in the NVE ensemble was carried out to remove any residual non-equilibrium effects and stabilize the total energy.

To probe thermal transport between the polymer and water, the temperature of the PEG chain was instantaneously raised by 30 K under an NVT ensemble, while the surrounding water molecules were maintained at the previously equilibrated reference temperature using an independent NVT thermostat. This procedure corresponds to reassigning the atomic velocities of PEG atoms according to the Maxwell-Boltzmann distribution at the elevated temperature before the NVE simulation. The imposed temperature difference created a well-defined thermal gradient between PEG and water, initiating the relaxation process.

Finally, the system evolved under the micro-canonical (NVE) ensemble for 30 ps to monitor thermal relaxation dynamics in the absence of external energy exchange. All simulations were performed with a time step of 2 fs, and the SHAKE algorithm was applied to constrain all bonds involving hydrogen atoms.

To improve statistical reliability, ten independent simulations with different initial atomic velocities were performed for each condition analyzed. The reported averages and error bars correspond to mean ±5% deviations. This simulation framework provides a robust and reproducible protocol for analyzing nanoscale interfacial heat transfer and coupling mechanisms between polymer and solvent phases during thermal relaxation.

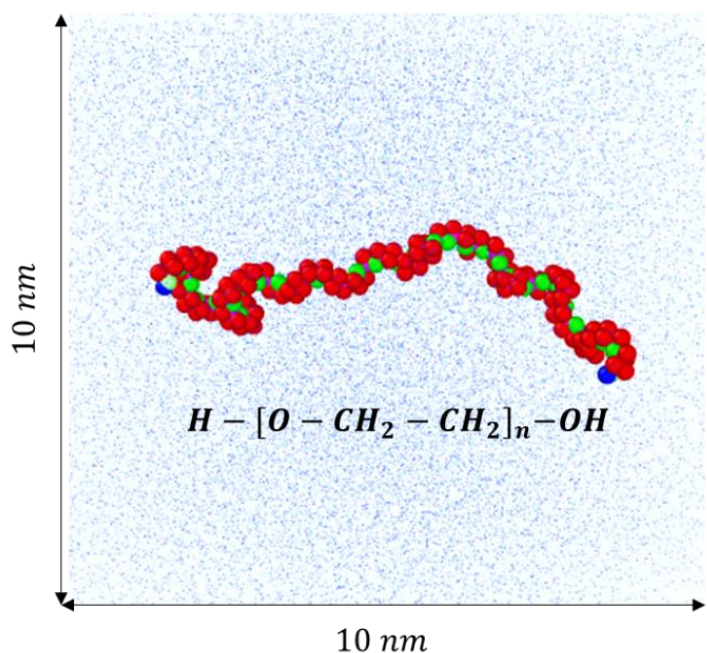

*Figure 1. Front view of the 3D simulation box ($10 \times 10 \times 10\ nm^3$) containing a 36-mer Polyethylene glycol (PEG) chain in water modeled explicitly. PEG atoms are shown as red (C), green (O), blue (H); water molecules are represented by light blue dots.*

## 3. Results and Discussion

### 3.1 Gyration Radius of PEG

In order to validate the reliability of our molecular dynamics simulations, we examined the temperature-dependent conformational behavior of a polyethylene glycol (PEG) 36-mer chain in aqueous solution by evaluating its radius of gyration (Rg) over the temperature range of 280 K to 370 K. As illustrated in Figure 2, the computed Rg decreases progressively with increasing temperature, indicating a gradual conformational compaction of the polymer chain.

This trend is consistent with the result of Chudoba et al [29], who employed a coarse-grained implicit solvent model over a broader temperature range (270–450 K) and reported a similar monotonic decrease of Rg with temperature, attributed to folding-like behavior of PEG chains at elevated temperatures. This phenomenon arises from reduced polymer–solvent interactions and weakened hydrogen bonding, leading to a more compact structure [37].

The agreement between our fully atomistic simulations and the referenced coarse-grained model supports the robustness of the CHARMM36 force field in describing PEG's temperature-dependent conformational response. Our results thus confirm the structural sensitivity of PEG to thermal conditions while providing atomistic insights into folding mechanisms in aqueous environments. As also noted by Chudoba et al. [29], shorter PEG chains such as the 36-mer exhibit less pronounced folding due to limited chain flexibility and lower entropic driving force. Hence, the observed compaction can be regarded as the onset of folding behavior that becomes more evident for longer chains, further validating the physical consistency of our findings.

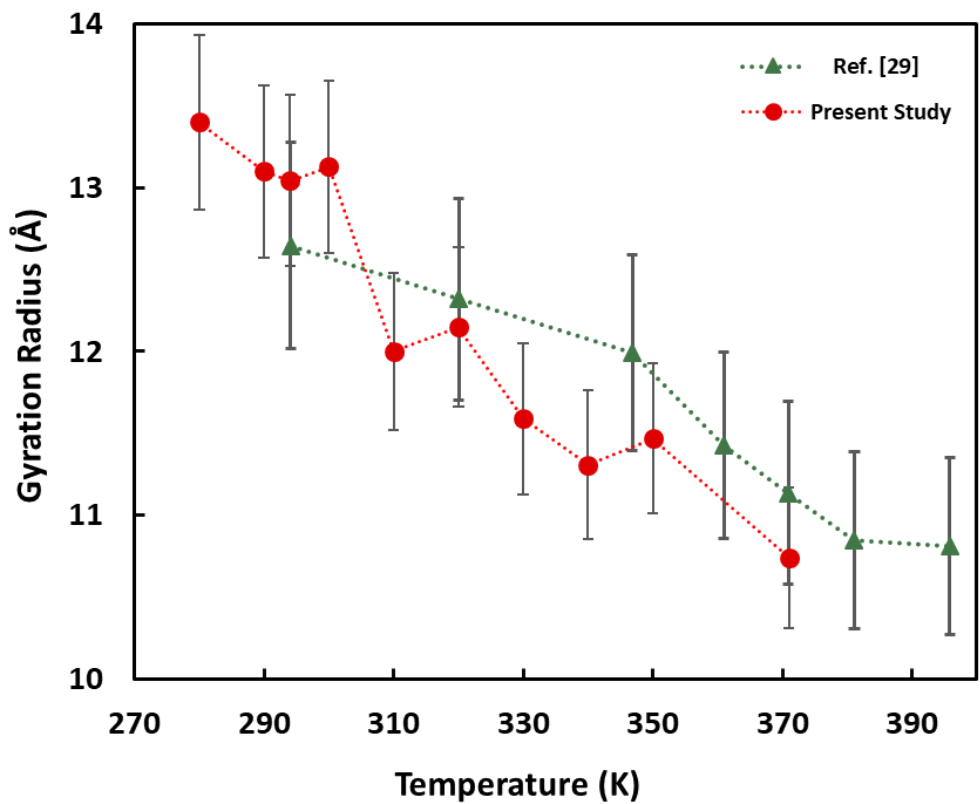

Figure 2. Radius of gyration of PEG (36-mer) vs. temperature. Simulation results (circles) are compared with the result of a coarse grained model (triangles)

To further put the results in context, comparison with experimental scaling data [38], shows that the $Rg$ of PEG in water at 50°C follows the empirical relation $Rg = 0.0215\, M_w^{0.583 \pm 0.031}$. For a 36-mer PEG chain (molecular weight ≈1602 g/mol), this predicts $R_g \approx 15$ Å, which aligns well with our simulation. Although, this empirical relation was derived for longer PEG chains, the agreement supports the experimental relevance of the present atomistic model.

## 3.2 Effect of PEG-Water Interaction Strength on Chain Structure

To explore the influence of molecular interactions on the structural and thermal behavior of PEG in water, we introduced a dimensionless scaling parameter $\chi = \varepsilon/\varepsilon_0$ by modifying the pairwise Lennard-Jones interaction strength between PEG and water molecules. Specifically, ε was scaled relative to its reference value ($\varepsilon_0$), considering three distinct interaction strengths: $\chi$ = 0.5, 1.0, and 2.0.

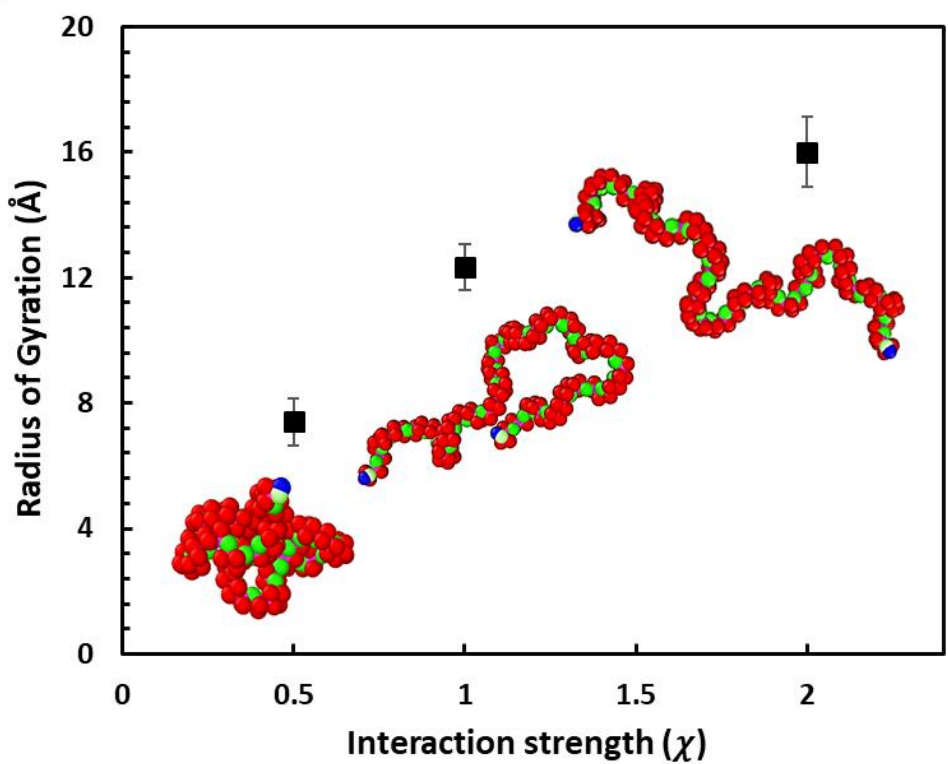

*Figure 3. Effect of the PEG–water interaction strength ($\chi$) on the radius of gyration of the PEG chain. All the simulations have been performed at 300 K.*

As shown in Figure 3, changing $\chi$ significantly affects the PEG conformation. At $\chi$ = 0.5, the chain adopts a compact, folded structure, indicating weakened interactions with water and reduced hydration. For $\chi$ = 1.0, the chain maintains its equilibrium configuration, while for $\chi = 2.0$, it becomes extended and unfolded structure, reflecting stronger PEG–water affinity.

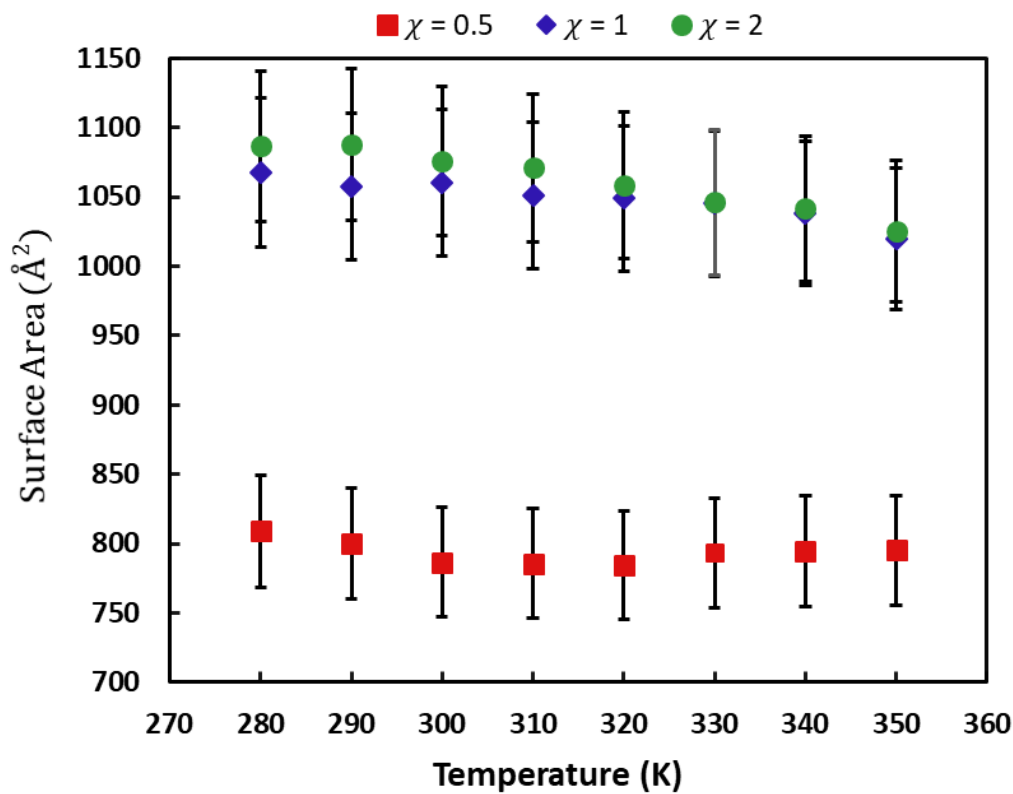

*Figure 4. Variation of 36-mer PEG–water interfacial area with temperature for different values of the Interaction strength ($\chi$)*

These conformational changes directly influence the interfacial area between PEG and water, as illustrated in Figure 3. The folded chain at $\chi = 0.5$ presents a smaller contact surface, while the extended one ($\chi = 2.0$) exhibits a considerably larger solvent-accessible interface as shown in figure 4. This confirms that both temperature and interatomic interaction strength ($\chi$) are critical parameters governing the structural flexibility of PEG and its effective contact area with the surrounding water environment.

## 3.3 Interfacial Thermal Conductance

Following the validation of the conformational behavior of our polymer model, the interfacial thermal conductance (h) between the PEG 36-mer chain and surrounding water was calculated using the transient non-equilibrium molecular dynamics (TNEMD) method [39]. This approach enables time-resolved monitoring of energy exchange between the polymer and solvent while providing an efficient route for extracting interfacial thermal transport coefficients.

After initial energy minimization, the system was equilibrated under an NPT ensemble for 1 ns at temperatures ranging from 280 K to 350 K. The simulation was then switched to an NVE ensemble for 50 ps to eliminate any residual non-equilibrium effects and stabilize the total energy. Subsequently, a heat pulse was imposed by instantaneously increasing the temperature of the PEG chain by 30 K in the NVT ensemble, while the surrounding water were maintained at the previously equilibrated reference temperature using an independent NVT thermostat during 50 ps. This procedure established a well-defined temperature difference between PEG and water, initiating the relaxation process. Finally, the system evolved under the NVE ensemble for 30 ps to capture the thermal relaxation dynamics in the absence of external energy exchange, as illustrated in figure 5.

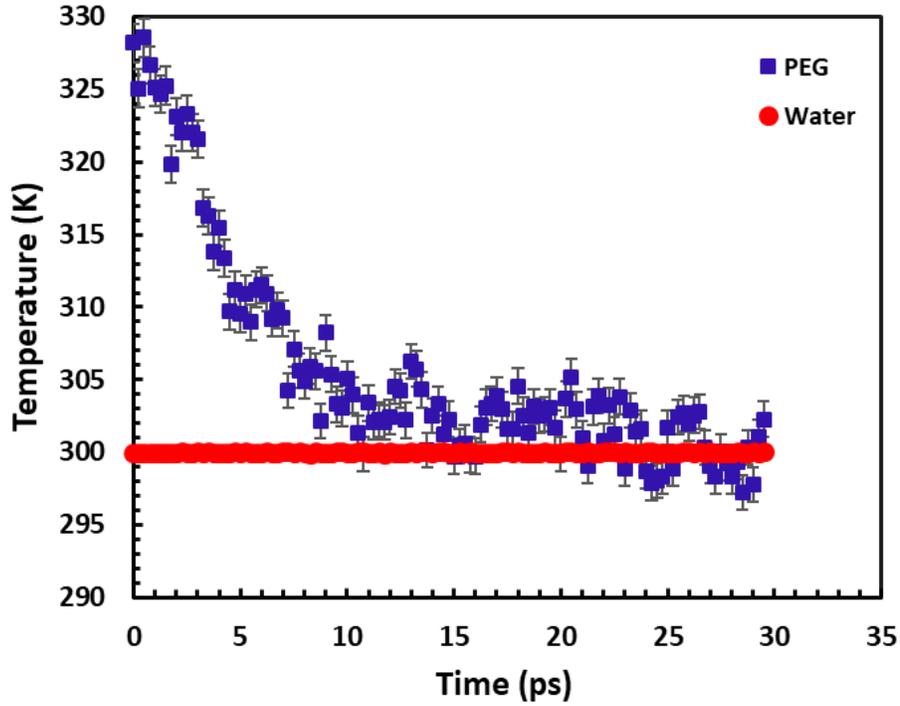

*Figure 5. Temperature relaxation of PEG (square) and water (circle) during 30 ps of thermal relaxation following a 30 K temperature jump applied to the PEG chain*

During this relaxation stage, the time-resolved values of PEG chain temperature ($T_P$), water temperature ($T_W$), and total energy of PEG ($E_t$), defined as the sum of kinetic and potential contributions were recorded. Averaging over ten independent trajectories provided smooth temperature and energy profiles with statistically consistent trends. The evolution of ($E_t$) as a function of the $\int \Delta T \, dt$ (time integral of the temperature difference), where $\Delta T = T_P - T_W$, showed a clear linear trend, validating the assumption of constant interfacial thermal conductance, as evidenced in Figure 6.

Based on the energy balance equation:

$$\frac{dE_t}{dt} = -hA\,(T_P - T_W) \tag{2}$$

and its integrated form:

$$E_t = -hA \int_0^\infty (T_P - T_W)\,dt + E_0 \tag{3}$$

the slope of the $E_t$ vs. $\int \Delta T\, dt$ curve yields the product hA [40], from which h is calculated using the solvent-accessible interfacial area (A) extracted from the simulation using OVITO software [41]. For each trajectory, the surface meshes were constructed around the PEG chain using the Construct Surface Mesh modifier with the Alpha Shape method, which accurately identifies the molecular boundary separating the polymer from the solvent. The obtained surface area values were averaged over ten independent simulations for each temperature to improve statistical accuracy. To improve the precision of (h), data from the initial non-linear stage and long-time fluctuations of the thermal relaxation were excluded from curve fitting process. The resulting values of h reflect the rate of thermal energy transfer across the PEG–water interface, capturing the temperature-dependent nature of interfacial thermal transport at the molecular scale.

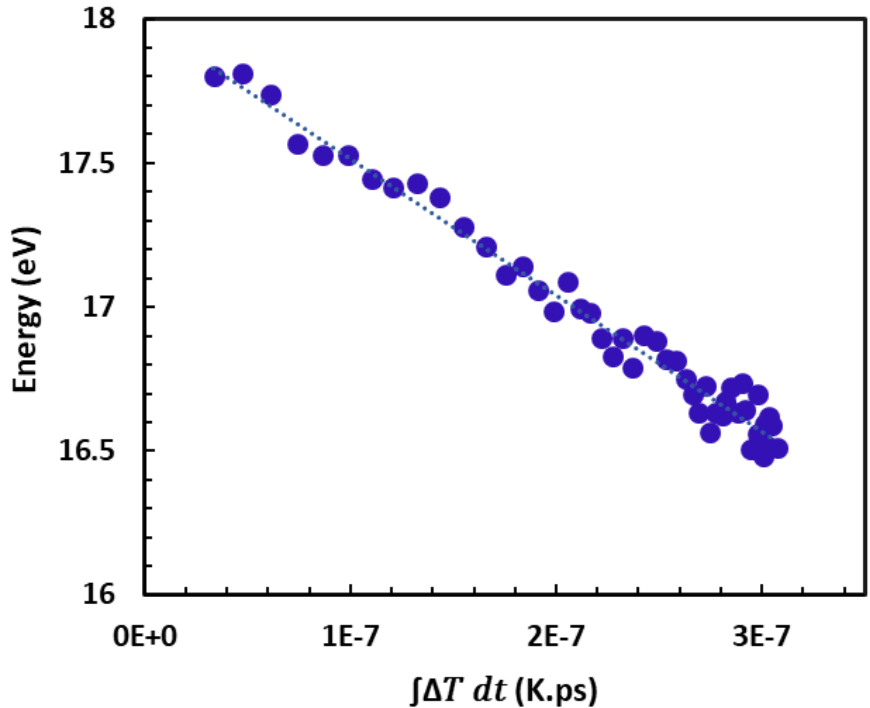

*Figure 6. Total energy of PEG molecule as a function of the time-integrated temperature difference ∫ΔT dt. The slope is used to compute the interface thermal conductance, see text for details.*

### 3.4 Estimation of the Interfacial Thermal Conductance (h)

To evaluate the interfacial thermal conductance (h) between the PEG chains and water, a series of molecular dynamics simulations were performed in which the PEG chain was first equilibrated at various temperatures ranging from 280 K to 350 K. For each equilibrated system, the radius of gyration and the interfacial contact area between PEG and water were calculated. In parallel, a thermal relaxation process was initiated by instantaneously increasing the temperature of the PEG chain, while monitoring the temporal evolution of the system energy as heat is transferred from PEG to the surrounding water.

According to Equation 3, $E_t = -hA \int_0^\infty (T_P - T_W)dt + E_0$, the interfacial thermal conductance (h) can be extracted by combining the slope of the energy decay curve (as shown in Figure 6) with the interfacial area (A) obtained from the OVITO analysis (figure 4).

Figure 7 presents the calculated values of h as a function of temperature for different interaction strengths ($\chi = 0.5, 1$ and $2$). For $\chi = 0.5$, the PEG chain adopts a compact conformation with a smaller interfacial area but a higher internal energy state, resulting in faster thermal relaxation and a relatively larger values of h [42]. By contrast, for $\chi = 1$ and $2$, the PEG chains remain extended with larger solvent-accessible area and lower internal energy, leading to slower relaxation dynamics and smaller values of the interfacial thermal conductance.

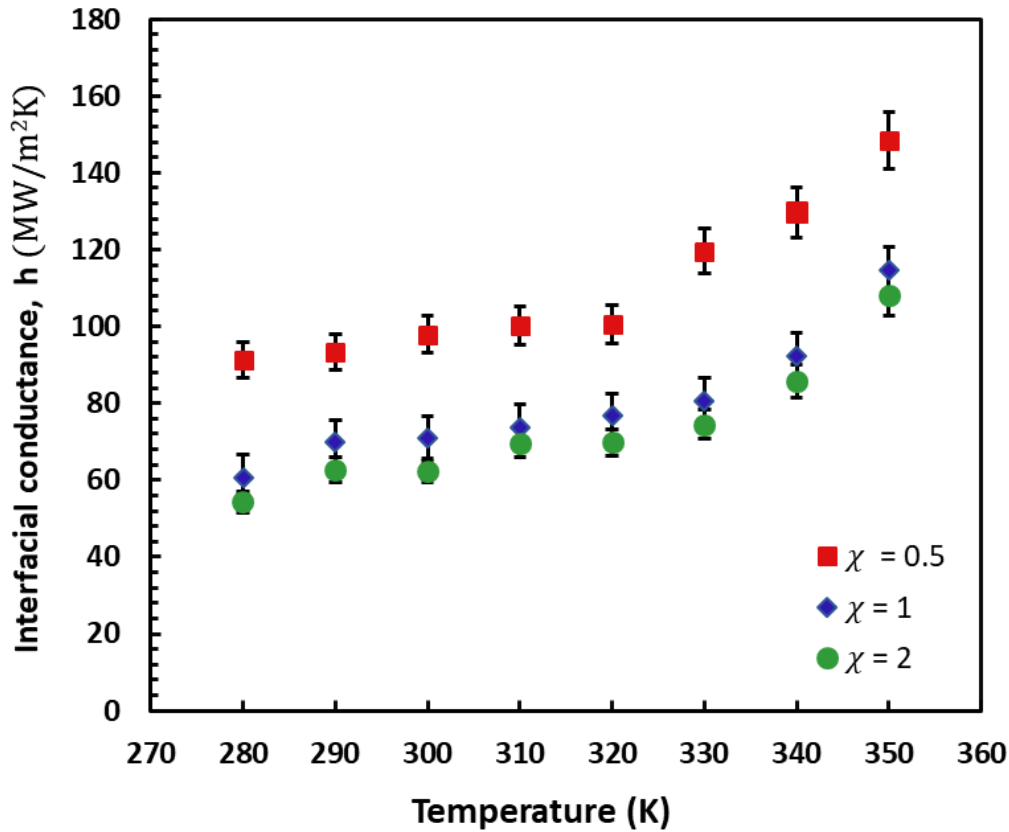

*Figure 7. Temperature-dependent interfacial thermal conductance (h) between PEG and water for different interaction strengths ($\chi$)*

These results underscore the dual influence of temperature and molecular interaction strength on interfacial thermal conductance h. Specifically, the observed trends indicate that (h) is highly sensitive to molecular-scale structural factors such as polymer conformation, interfacial surface area, and hydrogen-bonding strength. This finding provides a tunable framework for modulating the efficiency of heat transfer in PEG-based systems though controlled solvent quality or chemical functionalization.

## 3.5 Combined Effect of Temperature and Interaction Strength on the Total Interfacial Thermal Conductance (G)

To elucidate the combined impact of temperature and PEG–water interaction strength on interfacial heat transfer, we analyzed the total interfacial thermal conductance defined as $G = h \times A$, where h is the interfacial thermal conductance per unit area and A is the interfacial contact area.

As shown in Figure 8, G exhibits a consistently increase with temperature for all values of the coupling parameter $\chi$. This monotonic rise confirms that temperature is the primary factor governing interfacial energy transport. By contrast, the effect of $\chi$ is notably weaker: for any given temperature, the variation in G with different interaction strengths remains relatively small, indicating that molecular interaction strength does not directly control interfacial thermal conductance.

Instead, $\chi$ appears to influence G indirectly by altering the polymer conformation and interfacial area. The convergence of G at lower temperatures and their gradual increase at higher temperatures further highlights that the structural sensitivity to interaction strength becomes more pronounced under elevated thermal conditions. These observations suggest that, while the interaction strength contributes through secondary structural pathways, temperature remains the dominant driving force of interfacial thermal conductance.

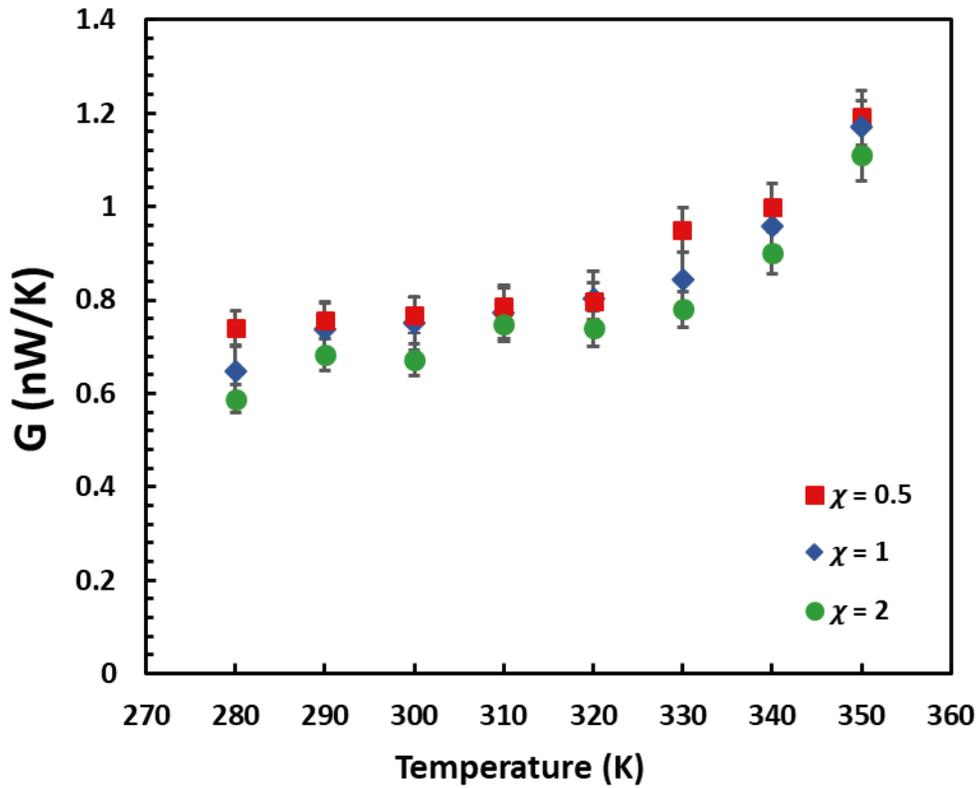

Figure 8. Temperature dependence of the total interfacial thermal conductance G for different PEG-water interaction strengths χ.

## 3.6 Vibrational Coupling Analysis Using VDOS Overlap

To further investigate the underlying microscopic mechanisms governing the variations in interfacial thermal conductance (h), we analyzed the vibrational density of states (VDOS) of both the PEG chain and the surrounding water molecules [43]. To compute the vibrational density of states (VDOS), we first extracted the velocity autocorrelation functions (VACF) separately for the PEG chain and surrounding water molecules based on atom types using MD trajectory outputs [44, 45]. The VACF data were then processed using SciDAVis [46], where we applied Fast Fourier Transform (FFT) to convert the time-domain signals into frequency-domain spectra, yielding the VDOS profiles. Curve fitting was performed to smooth out the spectral features. Finally, the spectral overlap between PEG and water was quantitatively evaluated in MATLAB by integrating the normalized VDOS curves under various temperature and interaction strength conditions. VDOS spectra were computed at different temperatures and PEG–water interaction strengths ($\chi$) as represented in Figure 9. The spectral overlap between the two subsystems was quantified using a frequency-based overlap integral defined by:

$$S = \frac{\int DoS_{PEG}(\omega) DoS_{Water}(\omega) d\omega}{\left(\sqrt{\int DoS_{PEG}^2(\omega) d\omega}\right)\left(\sqrt{\int DoS_{Water}^2(\omega) d\omega}\right)} \qquad (4)$$

where $\omega$ is the angular frequency [47]. Despite changes in temperature and interaction strength, the computed VDOS overlaps remain relatively constant across conditions as shown in table 1. Specifically, at a fixed value of temperature, varying $\chi$ has only a minor effect on the VDOS overlap. This suggests that the direct vibrational coupling between PEG and water remains largely unchanged with respect to the interfacial interaction strength.

Moreover, while increasing temperature introduces slight variations in the VDOS overlap, these changes are not significant enough to account for the effect of temperature pronounced differences observed in the interfacial thermal conductance h) and the combined conductance $G = h \times A$ analyzed previously.

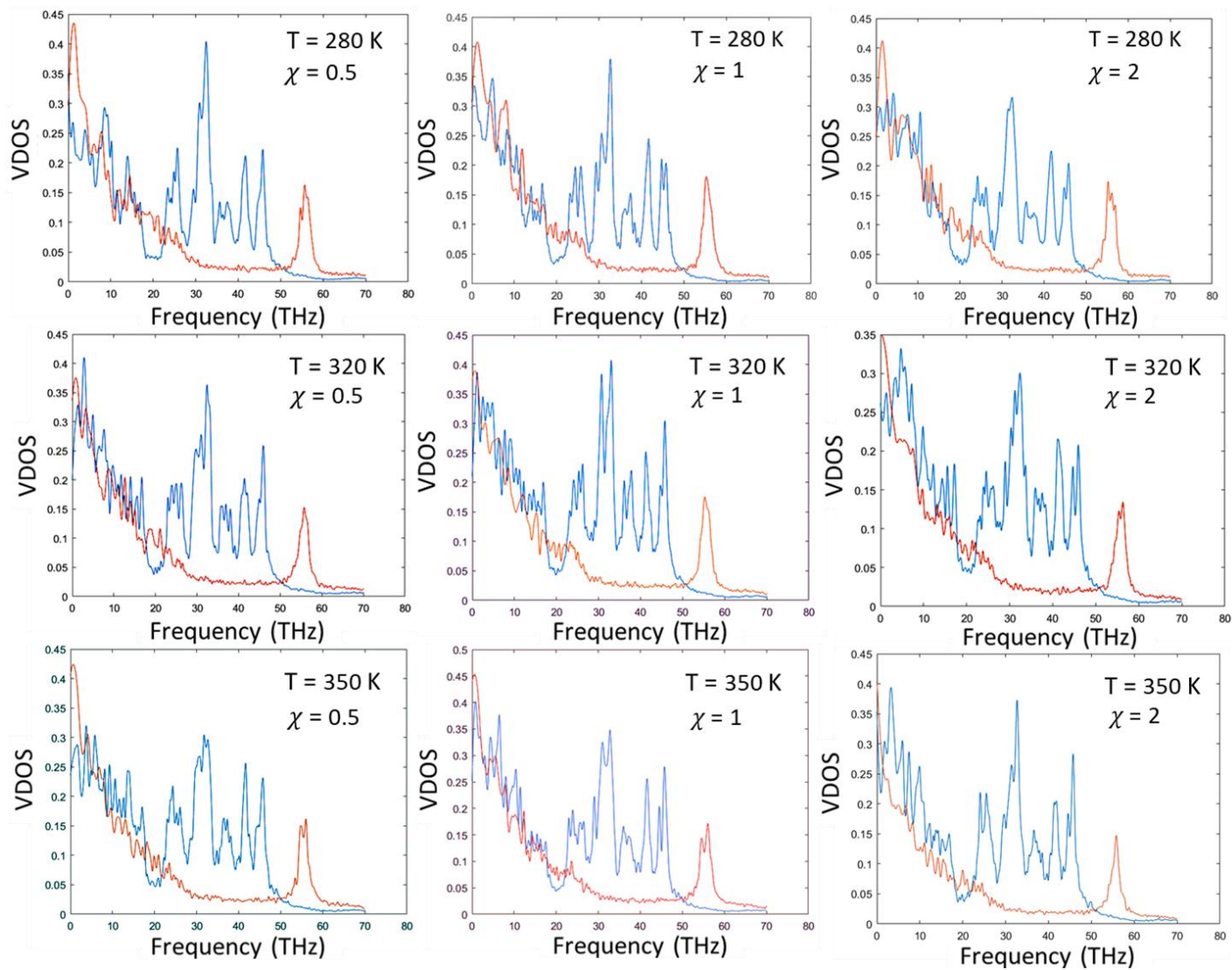

Figure 9. Vibrational density of states (VDOS) of PEG (blue) and water (red) at three temperatures (280 K, 320 K, and 350 K) and for the interaction parameters, χ=0.5,1.0,2.0.

| Temperature(K) | χ = 0.5 | χ = 1 | χ = 2 |
|---|---|---|---|
| 280 | 0.758 | 0.813 | 0.826 |
| 320 | 0.806 | 0.795 | 0.811 |
| 350 | 0.786 | 0.820 | 0.805 |

Table 1. Spectral overlap between the vibrational density of states (VDOS) of PEG and water at various temperatures and interaction strengths (χ). The computed spectral overlaps range from 0.758 to 0.826, and show only minor variation across conditions.

These findings support the conclusion that the dominant factors affecting interfacial heat transfer are structural primarily the polymer conformation and the extent of the interfacial area rather than changes in the vibrational spectral matching. In essence, stronger or weaker molecular interactions modify heat transfer efficiency not by affecting phonon coupling, but by reshaping the polymer interfacial geometry thus altering the energy relaxation pathway.

### 3.7 Potential Energy Analysis

Now that we have seen that changes in phonon overlap is practically independent of the value of the $\chi$ parameter, we seek to interpret the relative insensitivity of $\chi$ on the interface thermal conductance as seen in Figure 8. In particular, we see that despite lower interaction strength, the chains interacting weakly with water are characterized by a total thermal conductance close to that of interacting polymer chains. This effect should be related to the different conformation of the weakly interacting chain, as illustrated in Figure 3. In particular, for $\chi=0.5$, the polymer chain adopts a more compact structure, as compared to the extended structures adopted by the polymer chain for $\chi>1$. This change in conformation may lead to different coordination of the PEG atoms thus impacting heat transfer to water. To quantify this effect, we have calculated in Table II the coordination number of a PEG oxygen atom. To define the coordination, we have computed the average number of water oxygen atoms located within a 3.5 Å cutoff from the PEG oxygen atoms. Based on Table II, we conclude that polymer chains interacting strongly with water maximize the number of contacts between the chain and water. Conversely, the polymer chain with χ=0.5 has a relatively low coordination number.

| Total PEG-water O-contacts (≤ 3.5 Å) | χ = 0.5 | χ = 1 | χ = 2 |
|---|---|---|---|
| | 40.997 | 44.607 | 52.11 |

Table2. Time-averaged number of O (water) atoms located within 3.5 Å of O (PEG) atoms for different interaction strengths χ.

To further characterize the local structure around the polymer chains, Figure 10 a) displays the pair distribution function between PEG oxygen atoms and water oxygen atoms for the different values of χ. No significant difference is seen at short distances. The only noticeable differences concern the amplitudes of the secondary and tertiary peaks, located respectively at distance ~4.5 and 8 Angstroms, which amplitudes increase with the interaction parameter χ. We have also computed the corresponding distribution between, respectively PEG C atoms and water oxygen atoms and between PEG H atoms and water oxygen atoms and found no significant difference when the value of χ has been changed. Therefore, we conclude that the relatively high value of thermal conductance for χ=0.5 cannot be explained by an increasing number of PEG-water interacting atoms. To go further, we have computed the total internal energy of the polymer chains before the application of a heat pulse. The results displayed in Figure 10 b) show that the weakly interacting chain (χ=0.5) has a relatively high internal energy, probably due to the compact structure of the chain, see Figure 3. As a result, this compact configuration may store more internal energy that can restore more energy to surrounding water, explaining the relative enhanced heat transfer despite weaker direct interactions with water.

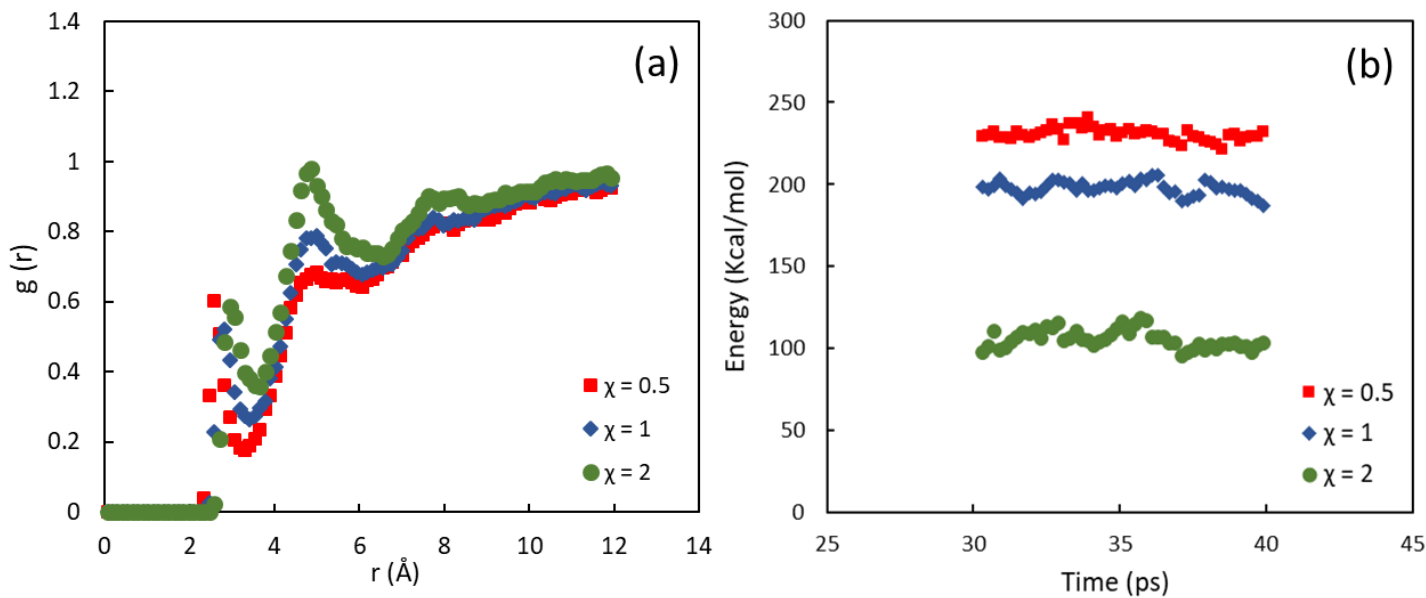

*Figure 10a: pair distribution function between the PEG oxygen atoms and water oxygen atoms for the different values of χ. 10b: internal energy of the PEG chains before the application of a heat pulse.*

### 3.8 Analysis of the Chain Mobility via Mean Square Displacement (MSD)

To gain deeper insight into the temperature-dependent enhancement of interfacial thermal conductance, the molecular mobility of the PEG chain and surrounding water molecules was analyzed through the time evolution of their mean square displacement (MSD). Figure 11 presents the time evolution of MSD for PEG and water for $\chi = 0.5$, 1 and 2.

The diffusion coefficients (D) were obtained from the linear regime of the MSD-time curves using the Einstein relation, $D = \frac{1}{6}\frac{d\langle r^2(t)\rangle}{dt}$ where the slope represents the rate of molecular diffusion in three dimensions [48]. The results show that both PEG and water exhibit a steady increase in D with temperature, indicating enhanced molecular motion and faster diffusion at elevated thermal energies. The diffusion coefficient of water molecules is significantly larger than that of PEG, consistent with the higher mobility of the solvent phase [49].

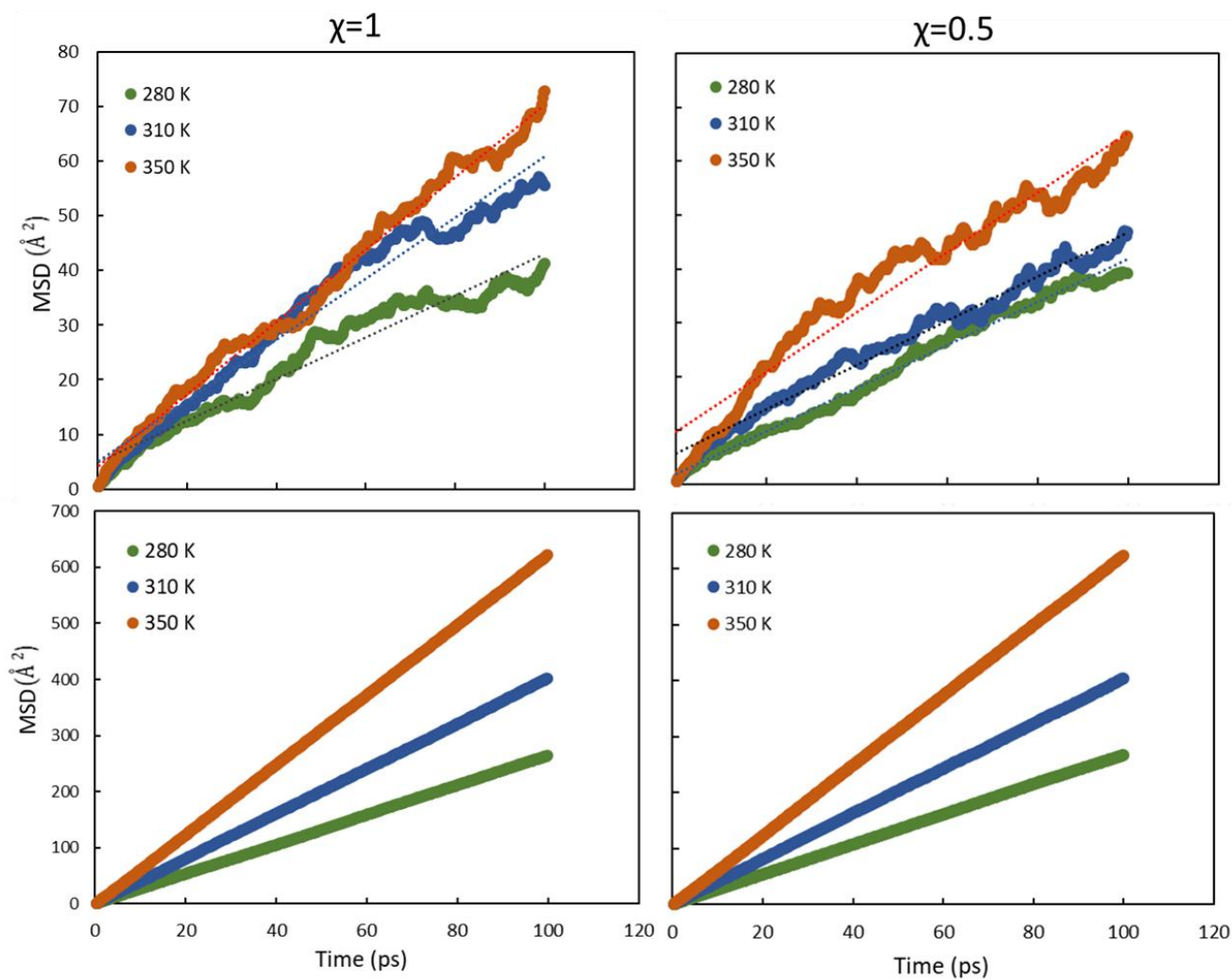

*Figure 11. Time-dependent mean square displacement (MSD) of PEG (top) and water molecules (bottom) at different temperatures and interaction strengths (χ = 0.5, 1).*

The influence of the PEG-water interaction strength ($\chi$) on the MSD slope is comparatively weak, suggesting that molecular interaction strength plays a secondary role relative to temperature in determining dynamic mobility. This observation aligns with the trends observed in interfacial thermal conductance, where temperature was identified as the dominant factor controlling heat transfer efficiency.

The enhanced molecular mobility at higher temperatures provides a mechanistic explanation for the increased interfacial thermal conductance observed in the relaxation phase. Greater diffusivity and dynamic fluctuations promote more efficient energy redistribution across the PEG-water interface, consistent with simulation studies on thermal transport at solid-liquid interfaces [49, 50]. In particular, in [50], it is shown that molecular diffusivity enhances inelastic heat transfer at solid-liquid interfaces. Although this latter study deal with solid-water interfaces, it is reasonable to posit that similar qualitative conclusions apply to the system considered here, due to the relatively lower diffusivity of the polymer chains as compared to water.

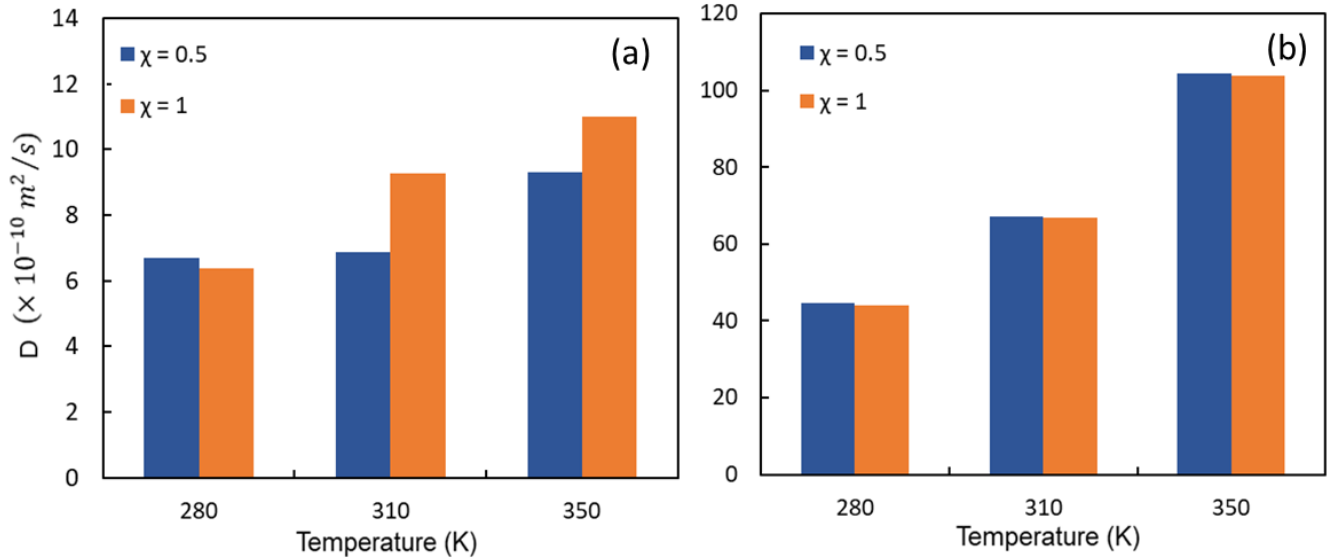

*Figure 12. Diffusion coefficient (D) as a function of temperature for (a) PEG and (b) water for different interaction strengths ($\chi$ = 0.5, 1). The values were obtained from the long-time behavior of the MSD curves and the Einstein relation.*

As shown in Figure 12, the diffusion coefficients of both PEG and water increase systematically with temperature. The weak dependence of D on the interaction strength $\chi$ further supports the conclusion that temperature -rather than polymer-water affinity- is the primary factor governing molecular mobility in the system and subsequently interfacial heat transfer. The monotonic rise in diffusivity at elevated temperatures provides direct dynamical evidence for the enhanced energy redistribution capability of the polymer, offering a complementary explanation for the temperature-driven increase in interfacial thermal conductance observed in this study.

## 4. Conclusion

In this work, atomistic molecular dynamics simulations were conducted to investigate the interfacial thermal conductance (h) between a PEG 36-mer polyethylene glycol (PEG) chain and water. The results confirm that heat transfer across the polymer-water interface is strongly governed by both temperature and interfacial interaction strength, with temperature emerging as the dominant factor controlling thermal transport. Structural analysis based on the radius of gyration and solvent-accessible surface area revealed that variations in the interaction strength primarily affect interfacial thermal conductance indirectly by altering polymer conformation and its effective contact geometry with water.

Systems with more compact PEG conformations ($\chi = 0.5$) exhibit smaller interfacial surface areas but store higher internal energy after the heat pulse, behaving as stronger thermal sources during relaxation. This behavior reflects higher potential energy storage and an effectively greater heat capacity within the compacted PEG chain [51]. Conversely, at higher interaction strengths ($\chi = 2$), extended PEG conformations with larger contact areas but lower initial internal energy resulted in reduced interfacial heat transfer.

The vibrational density of states (VDOS) analysis indicated that the spectral overlap between PEG and water remains nearly constant across all the conditions investigated, suggesting that phonon coupling is not the main mechanism explaining the observed variation in h. Instead, the results indicate an internal-energy-driven mechanism, where the PEG chain's capacity to absorb and release energy plays a decisive role in modulating interfacial thermal conductance.

Additionally, the analysis of the mean square displacement (MSD) demonstrated that both PEG and water exhibit enhanced molecular mobility with increasing temperature, reinforcing the link between dynamic motion and interfacial energy exchange. This correlation is in line with previous simulation studies that associate higher diffusivity with increased interfacial heat flux at liquid-solid interfaces[49, 50].

Overall, these findings provide a molecular-level picture of how polymer conformation, interaction strength, and molecular diffusion govern interfacial thermal conductance in polymer/water systems. The insights presented here offer practical guidelines for tuning the thermal transport properties of PEG-based systems through chemical modification or solvent engineering, with potential implications for drug delivery, nanoparticle-based cancer therapy and energy conversion nanofluidic devices.